\documentclass[twocolumn,pra,amsmath,amssymb,superscriptaddress,showpacs]{revtex4-1}

\usepackage{graphicx}
\usepackage{dcolumn}
\usepackage{bm}

\usepackage{color}
\newcommand{\ket}[1]{\left\vert #1\right\rangle}
\newcommand{\bra}[1]{\left\langle #1\right\vert}

\newcommand{\beq}{\begin{eqnarray}}
\newcommand{\eeq}{\end{eqnarray}}

\begin{document}

\title{Environmental dynamics, correlations, and the emergence of noncanonical equilibrium states in open quantum systems}
\author{Jake Iles-Smith}
\email{jakeilessmith@gmail.co.uk}
\affiliation{{Photon Science Institute \& School of Physics and Astronomy, The University of Manchester, Oxford Road, Manchester M13 9PL, United Kingdom}}
\affiliation{Controlled Quantum Dynamics Theory, Imperial College London, London SW7 2AZ, United Kingdom}
\author{Neill Lambert}
\affiliation{QCMRG, CEMS, RIKEN, Saitama, 351-0198, Japan}
\author{Ahsan Nazir}%
\email{ahsan.nazir@manchester.ac.uk}%
\affiliation{{Photon Science Institute \& School of Physics and Astronomy, The University of Manchester, Oxford Road, Manchester M13 9PL, United Kingdom}}
\affiliation{Controlled Quantum Dynamics Theory, Imperial College London, London SW7 2AZ, United Kingdom}
\date{\today}

\begin{abstract}

Quantum systems are invariably open, evolving under surrounding influences rather than in isolation. Standard open quantum system methods eliminate all information on the environmental state to yield a tractable description of the system dynamics. By incorporating a collective coordinate of the environment into the system Hamiltonian, we circumvent this limitation. Our theory provides straightforward access to important environmental properties that would otherwise be obscured, allowing us to quantify the evolving system-environment correlations. As a direct result, we show 
that the generation of robust system-environment correlations that persist into equilibrium (heralded also by the emergence of non-Gaussian environmental states) renders the canonical system steady-state 
almost always incorrect. The resulting equilibrium states deviate markedly from those predicted by standard perturbative techniques
and are instead fully characterised by thermal states of the mapped system-collective coordinate Hamiltonian. We outline how noncanonical system states could be investigated experimentally to study deviations from canonical thermodynamics, with direct relevance to molecular and solid-state nanosystems.

 \end{abstract}
\maketitle

\section{Introduction}
A quantum system coupled to its macroscopic
environment constitutes
a challenging theoretical problem in which the large number of
environmental
degrees of freedom can lead to both conceptual and practical
difficulties~\cite{leggett87,weissbook,nitzanbook,maykuhnbook,breuer2007theory}. The master equation formalism
has been developed to offer a
simple and intuitive approach for describing
such systems; the complex dynamical
evolution of
the many-body environment
is not tracked explicitly, but 
instead only its
effect on the reduced state of
the system of interest is considered, 
eliminating all information 
on the environmental state.
In most cases one must
also rely on a series
of assumptions
in deriving a tractable master equation~\cite{Note1}. 
These customarily neglect the formation of system-bath correlations and
lead to the eventual thermalisation of the system with respect
solely to its internal Hamiltonian, resulting in canonical equilibrium 
states.

Here, by incorporating a collective coordinate of the environment into an effective system Hamiltonian, we develop a master equation formalism that can overcome such restrictions. This enables us to straightforwardly determine key environmental properties 
as well as track the dynamic generation of correlations between the system and bath. Specifically, we characterise the departure 
of the environment 
from its initial Gaussian thermal state due to interactions 
with the quantum system, and show that the resulting correlations---measured in terms of the mutual information---can have a profound effect on the system dynamics too, even persisting into the steady-state. 
We demonstrate that system-bath correlations are in fact generated on two distinct timescales, with
long-lived 
correlations leading to a departure of the system steady-state from canonical equilibrium, as would otherwise 
be expected from a perturbative (Born-Markov) treatment, 
thus heralding the failure of the accepted statistical mechanics view of thermalisation~\cite{breuer2007theory}. 
Correctly capturing system-environment correlations is 
hence 
shown to be crucial in order to properly describe 
both the system transient and equilibrium behaviour. 
As a further, and unique, aspect of our approach we illustrate how noncanonical equilibrium 
states can still 
be characterised in terms of 
thermal states, but now with respect to the {\it effective} system-collective coordinate Hamiltonian. 
This also reveals
simple experimental signatures by which deviations from canonical thermodynamics can be observed in real physical systems, for example through measurements of system populations.

\section{Reaction coordinate mapping and master equation}

Our method relies on keeping track of a collective environmental degree of freedom, and to do so
we 
make use of the 
reaction coordinate mapping~\cite{garg:4491,thoss:2991,cao97,hughes:124108,hartmann:11159,Roden,martinazzo:011101,Woods:arXiv1111.5262,imamoglu94,garraway97}; we take a quantum 
system 
coupled to a bosonic environment and map to a model in which a
collective mode of the environment, known as the reaction
coordinate (RC), is incorporated within an effective system
Hamiltonian. 
We then treat the residual environment within a full second-order
Born-Markov master equation formalism. By comparing the RC
master equation to the numerically exact hierarchical equations of motion (HEOM)~\citep{Tanimura2,Tanimura3}
we demonstrate essentially perfect agreement in the dynamics 
across all timescales (see below). Thus, all
important system-bath, and indeed intra-bath~\cite{bera13,bera14}, correlations are
incorporated into the system-RC 
Hamiltonian in the regimes we study.

Though our approach may be applied quite generally, 
we shall focus 
in this work 
on a two-level system (TLS) described by the spin-boson Hamiltonian~\cite{leggett87,weissbook,nitzanbook,breuer2007theory,maykuhnbook,makri95a,PhysRevLett.105.050404,alvermann2009,wang2008coherent,Anders07,winter09,silbey:2615,mccutcheon:114501,jang08,PhysRevLett.103.146404} (with $\hbar=1$): 
\begin{equation}\label{eq:SB}
H=\frac{\epsilon}{2} \sigma_{z}+\frac{\Delta}{2}\sigma_x+\sigma_z\sum\limits_k f_k(c_k^\dagger+c_k)+\sum\limits_k\nu_k c_k^\dagger c_k.
\end{equation} 
Here, $c_k^\dagger$ ($c_k$) are creation (annihilation)
operators for bosonic modes
of frequency $\nu_k$, which couple to the TLS with strength $f_k$, and $\sigma_i$ ($i=x,y,z$)
are TLS Pauli operators defined such that
$\sigma_z=|1\rangle\langle1|-|2\rangle\langle2|$. In the absence of the bath, the TLS splitting is determined by the bias $\epsilon$ and tunneling $\Delta$. The effect of the system-bath interaction can be completely
characterised by introducing the spectral
density~\cite{leggett87},
$J_{\rm SB}(\omega)=\sum_k f_k^2\delta(\omega-\omega_k)$.

\begin{figure}[t!]
\ \ \ \ \ \includegraphics[width=0.4\textwidth]{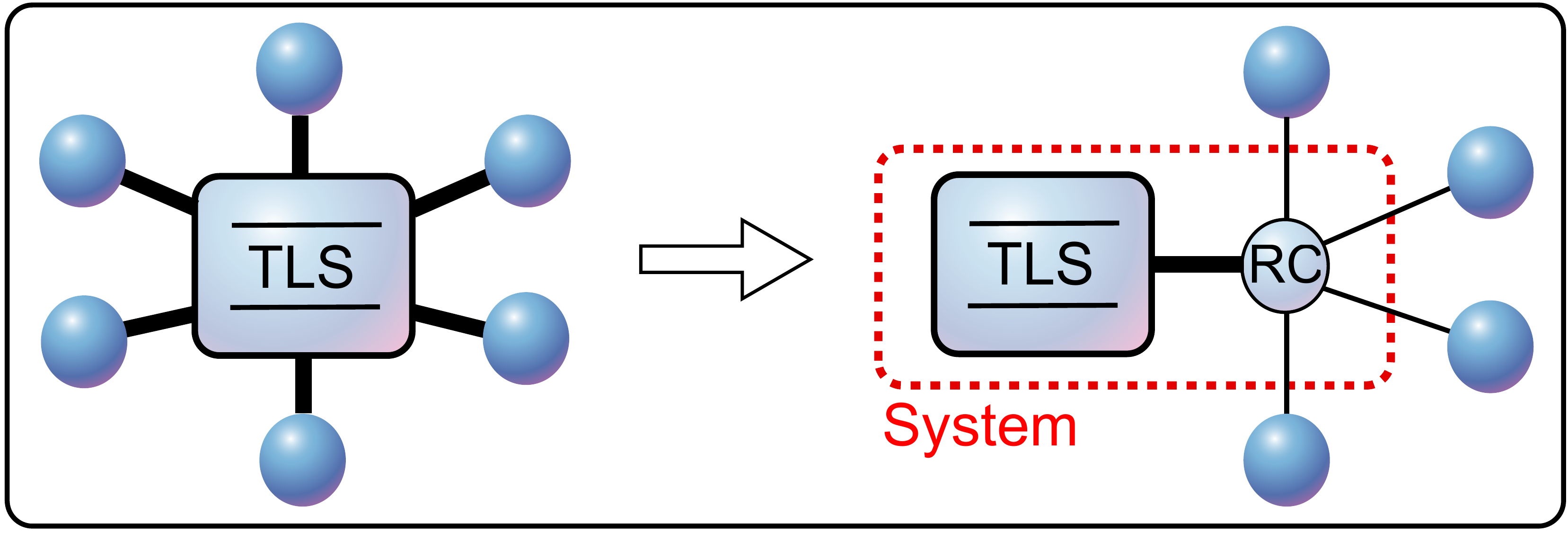}
\includegraphics[width=0.44\textwidth]{figure1b}
\caption{(Color online) 
Upper:~A TLS interacting with
a bosonic environment (left) is mapped to a TLS coupled
only to a collective mode,
which is in turn damped by a residual bath (right). Lower:~TLS population dynamics from the RCME (black, solid curves),
weak coupling theory (red, dash-dot curves), and converged HEOM data (points). Parameters: $\epsilon=0.5\Delta$,
$\omega_c=0.05\Delta$, $\beta\Delta=0.95$, and (a)
$\pi\alpha=0.1\Delta$, (b) $\pi\alpha=2.5\Delta$} \label{fig:pdyn}
\end{figure}

We now apply a normal mode transformation to Eq.~(\ref{eq:SB}) to
incorporate the most important environmental degrees of freedom into a new effective
system Hamiltonian. We carry out this procedure 
by first defining a collective coordinate of the environment~\cite{garg:4491}, the RC, which couples
directly to the TLS, and is in turn coupled to a residual harmonic
environment, as shown schematically in the upper panel of Fig~\ref{fig:pdyn}. This leads to a mapped Hamiltonian of the form:
\begin{align}\label{eq:RC}
H_{\rm RC}&=\frac{\epsilon}{2} \sigma_{z}+\frac{\Delta}{2}\sigma_x+\lambda\sigma_z(a^\dagger+a)+\Omega a^\dagger a+\sum\limits_k\omega_k b^\dagger_kb_k\nonumber\\
&+(a^\dagger+a)\sum\limits_k g_k(b_k^\dagger+b_k)+(a^\dagger+a)^2\sum\limits_k\frac{g_k^2}{\omega_k},
\end{align}
where the collective coordinate
is defined such
that:
\begin{equation}\label{eq:RCmode}
\lambda(\hat{a}^\dagger+\hat{a})=\sum\limits_kf_k(c_k^\dagger+c_k),
\end{equation}
with
coupling $\lambda^2=\sum_kf_k^2$ and frequency $\Omega=\lambda^{-1}\sqrt{\sum_k\nu_kf_k^2}$~\cite{thoss:2991,cao97,hughes:124108}.
The residual bath, denoted by creation (annihilation) operators $b_k^\dagger$ ($b_k$), now couples {\it only} to the RC
and is
characterised by an effective spectral density,
$J_{\rm RC}(\omega)=\sum_k g_k^2\delta(\omega-\omega_k)$. To describe
the action of the residual bath on the RC mode we need to relate this spectral density to the original spin-boson
$J_{\rm SB}(\omega)$. 
To do so, we follow Refs.~\cite{garg:4491,hughes:124108,PhysRevB.30.1208}
and replace the
TLS with a classical coordinate $q$ moving in a potential
$V(q)$.
As outlined 
in Appendix~\ref{RCmap}, 
by considering the Fourier transformed equations of motion
for $q$ both before and after the mapping,
we may relate $J_{\rm RC}(\omega)$ to the original $J_{\rm SB}(\omega)$.

To give a concrete example we now consider 
a Drude-Lorentz 
form 
commonly used to describe molecular systems~\cite{gilmore05}: 
$J_{\rm SB}(\omega)=\alpha\omega_c{\omega}/{(\omega^2+\omega_c^2)}$, 
with 
cut-off frequency $\omega_c$ and 
coupling strength $\alpha$. 
The mapping
we exploit
is not limited to this particular 
spectral density~\cite{Woods:arXiv1111.5262}, however it does allow us to
benchmark our results against data attained from the numerically
exact HEOM. 
By taking 
the RC spectral density
to be $J_{\rm RC}(\omega)=
\gamma\omega e^{-\omega/\Lambda}$,
we find that the equivalent spin-boson form is given by
$J_{\rm SB}(\omega)=4\gamma\omega\Omega^2\lambda^2/[(\Omega^2-\omega^2)^2+(2\pi\gamma\Omega\omega)^2]$ in the limit $\Lambda\rightarrow\infty$.
Hence, we
recover the Drude-Lorentz spectral density by identifying $\omega_c=\Omega/(2\pi\gamma)$, $\alpha=2\lambda^2/(\pi\Omega)$,
and choosing $\gamma$ such that $\omega_c\ll\Omega$~\cite{thoss:2991}.

Diagonalising the first line of Eq.~(\ref{eq:RC}) we can account for the original spin-boson interaction term in Eq.~(\ref{eq:SB}) to all orders, which ensures
that our formalism remains non-perturbative in the system-bath coupling strength $\alpha$.
We thus derive a
Born-Markov master equation for the reduced state of the composite TLS and RC,
$\rho(t)$, which captures their exact internal dynamics, while treating the coupling to the residual bath to
second order (see Appendix~\ref{sec:appendme}). 
This results in the
Schr\"odinger picture master
equation~\cite{breuer2007theory,PhysRevA.81.022117}:
\begin{align}\label{eq:full}
\frac{\partial\rho(t)}{\partial t}=-i\left[H_0,\rho(t)\right]-\left[\hat{A},\left[\hat{\chi},\rho(t)\right]\right]+\left[\hat{A},\left\{\hat\Xi,\rho(t)\right\}\right],
\end{align}
where we have assumed that only the {\it residual} environment remains in
thermal equilibrium throughout the evolution.
Here,
$\hat{A}=\hat{a}^\dagger+\hat{a}$,
\begin{align}
\hat{\chi}=&\gamma\int\limits_0^\infty d\tau\int\limits_0^\infty  d\omega\ \omega\cos(\omega\tau)\coth\left(\frac{\beta\omega}{2}\right) \hat{A}(-\tau),\label{eq:rate operators1}\\
\hat{\Xi}=&\gamma\int\limits_0^\infty d\tau\int\limits_0^\infty  d\omega \cos(\omega\tau)\left[H_0,\hat{A}(-\tau)\right],\label{eq:rate operators2}
\end{align}
with $\hat{A}(t)=e^{iH_0t}\hat{A}e^{-iH_0t}$ 
and $H_0=\frac{\epsilon}{2} \sigma_{z}+\frac{\Delta}{2}\sigma_x+\lambda\sigma_z(a^\dagger+a)+\Omega a^\dagger a$. To solve our master equation 
we truncate the RC space 
as necessary for 
convergence.

\section{Benchmarking}

In order to demonstrate the validity of the reaction coordinate master equation (RCME),
we benchmark its predictions for the TLS
population dynamics ($\rho_{11}=\langle1|\rho|1\rangle$)
against converged data
generated using the numerically exact HEOM
technique, see Appendix~\ref{heom} and Refs.~\onlinecite{Tanimura2,Tanimura3,ishizaki:234111}.
To give an illustrative example, in Fig.~\ref{fig:pdyn} we have taken parameters that are typical for excitonic energy transfer in molecular systems~\cite{ishizaki:234111,PhysRevLett.105.050404,Thorwart2009234,felix13}, with a representative value of $\Delta=200$~cm$^{-1}$ setting the other energy scales (i.e.~$t\sim1$~ps at $\Delta t=35$). The TLS is initialised in state $|1\rangle$, uncorrelated with both the RC and residual bath, which are taken to be in their respective thermal equilibrium states at a temperature $T=1/\beta$ ($\approx300$~K for $\Delta=200$~cm$^{-1}$). We observe essentially  
perfect agreement between the RCME and the
HEOM for a slow environment 
and
for 
spin-boson coupling strengths [(a) $\pi\alpha=20$~cm$^{-1}$, (b) $\pi\alpha=500$~cm$^{-1}$] encompassing the transition from the coherent to the incoherent regime~\cite{Note2}. 
In contrast, a standard 
weak coupling approach, treating the interaction term in Eq.~(\ref{eq:SB}) to second order, 
fails even qualitatively to capture the correct system behaviour for any 
parameters shown. 
It is thus clear that  environmental memory and
the generation of correlations
with the system---both of which are ignored in the weak coupling calculation---are crucial in order to
capture the correct
dynamical behaviour in this regime. 
Moreover, their combined impact
on the TLS can be accurately 
described simply through the TLS-RC coupling. This is somewhat remarkable, given that any non-thermal state effects of the original bath have been reduced solely 
to the action of a single mode on the TLS.

\section{Environmental dynamics and correlations}

The most important aspect of our formalism is that the inclusion of environmental degrees of freedom into the system Hamiltonian allows us to gain additional insight into the dynamics of both the environmental state and system-environment correlations. We do this by calculating two complementary measures; the RC non-Gaussianity~\cite{PhysRevA.78.060303,PhysRevA.82.052341}, which probes the environmental evolution:
\begin{equation}\label{eq:NG}
\delta_G[\rho_{\rm RC}(t)]=S(\varrho)-S\left(\rho_{\rm RC}(t)\right),
\end{equation}
and the TLS-RC quantum mutual information (QMI)~\cite{PhysRevA.72.032317}, characterising the 
correlations: 
\begin{equation}\label{eq:QMI}
\mathcal{I}(\rho_S:\rho_{\rm RC})=S(\rho_s)+S(\rho_{\rm RC})-S(\rho).
\end{equation}
Here, $\rho_{\rm RC}$ ($\rho_s$) is the reduced state of the RC (TLS) and $S(\chi)=-{\rm tr}\left(\chi\ln\chi\right)$ is the von-Neumann entropy.
The non-Gaussianity determines 
the distance 
from $\rho_{\rm RC}$ 
to the nearest Gaussian reference state $\varrho$, and is defined such that $\delta_G=0$ iff 
$\rho_{\rm RC}$ is Gaussian. 
The QMI quantifies the
total classical and quantum correlations shared between the TLS and RC ~\cite{PhysRevA.72.032317}.
As detailed in Appendix~\ref{QMInonGauss}, 
both measures 
act as rigorous lower bounds for the original spin-boson environment, which enables us to explore properties of the multi-mode 
bath and its correlations with the system simply through the single mode RC. 
Furthermore, in the limit that the Born approximation holds between the mapped system and residual environment, then the additive nature of the von-Neumann entropy implies that both these measures become {\it exact} for 
the original spin-boson environment.

\begin{figure}[t]
\center
\includegraphics[width=0.48\textwidth]{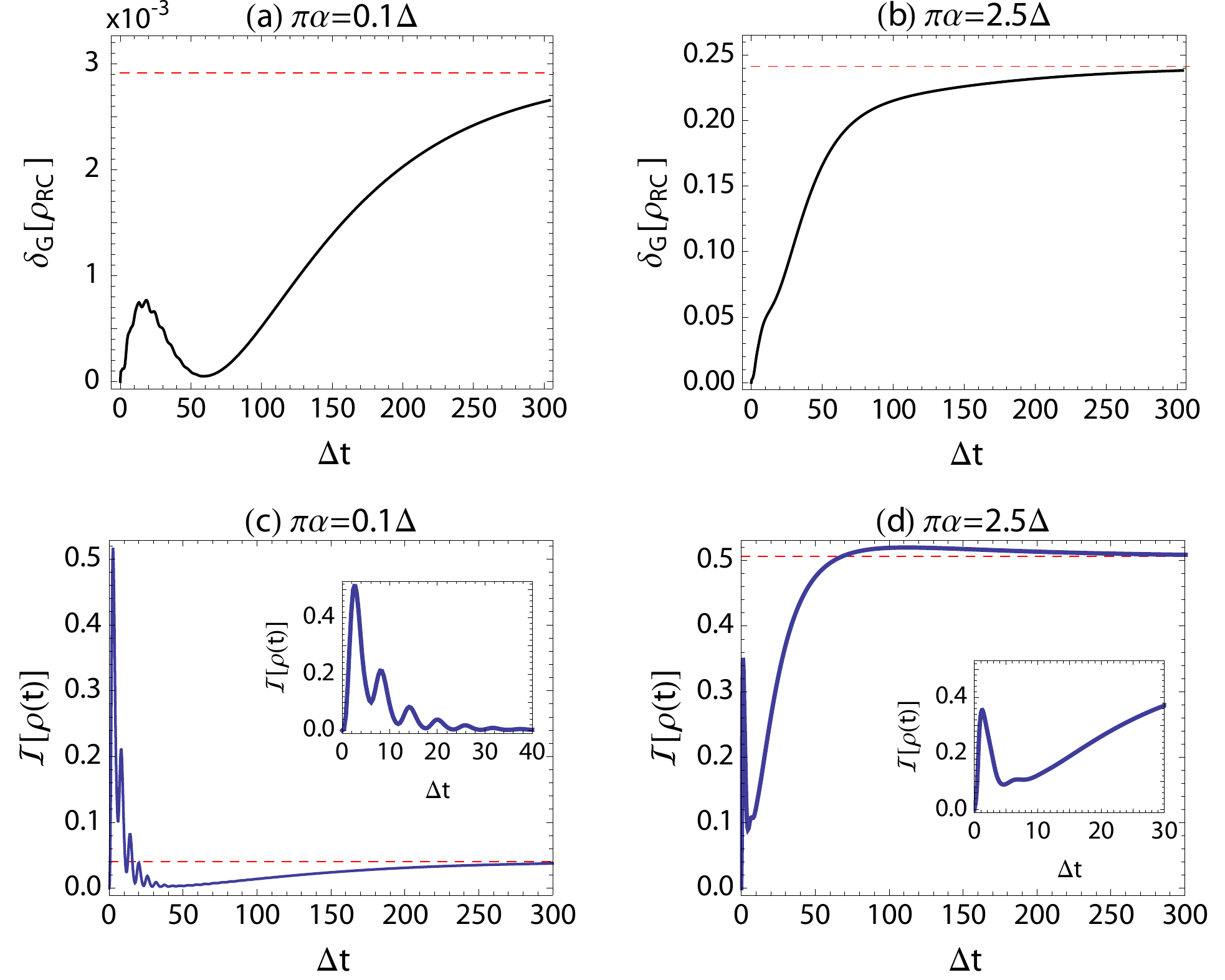}
\caption{(Color online) 
Dynamics of the RC non-Gaussianity (a,b) and TLS-RC mutual information (c,d) for weak and strong system-bath coupling strengths,
with steady state values indicated (dashed lines). 
Parameters: $\epsilon=0.5\Delta$,
$\omega_c=0.05\Delta$ and $\beta\Delta=0.95$, e.g.~for $\Delta=200$~cm$^{-1}$, $T\approx300$~K, and $t\sim10$~ps at $\Delta t=300$.}
\label{fig:NG}
\end{figure}

\begin{figure}[t]
\center
\includegraphics[width=0.235\textwidth]{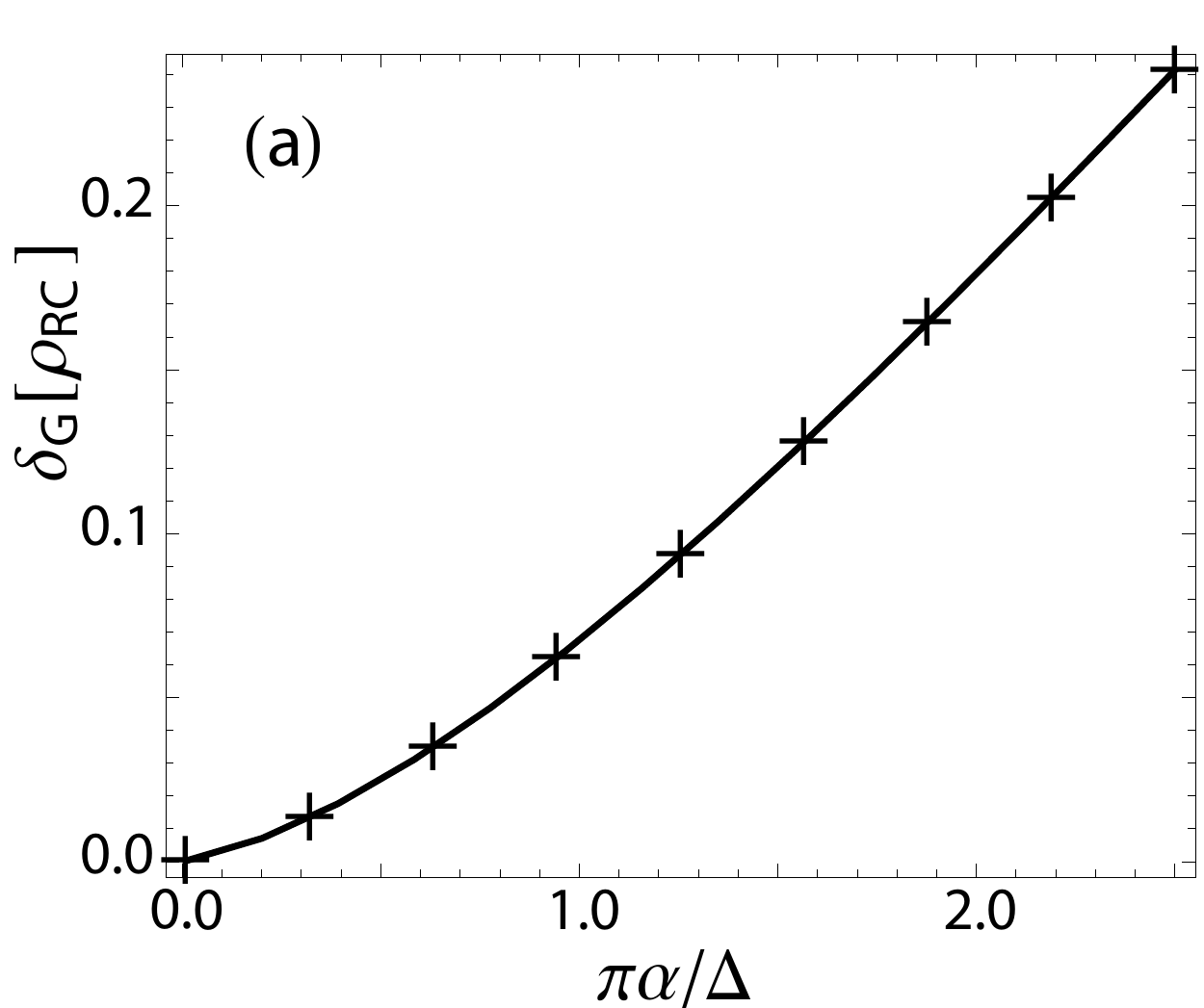}
\includegraphics[width=0.235\textwidth]{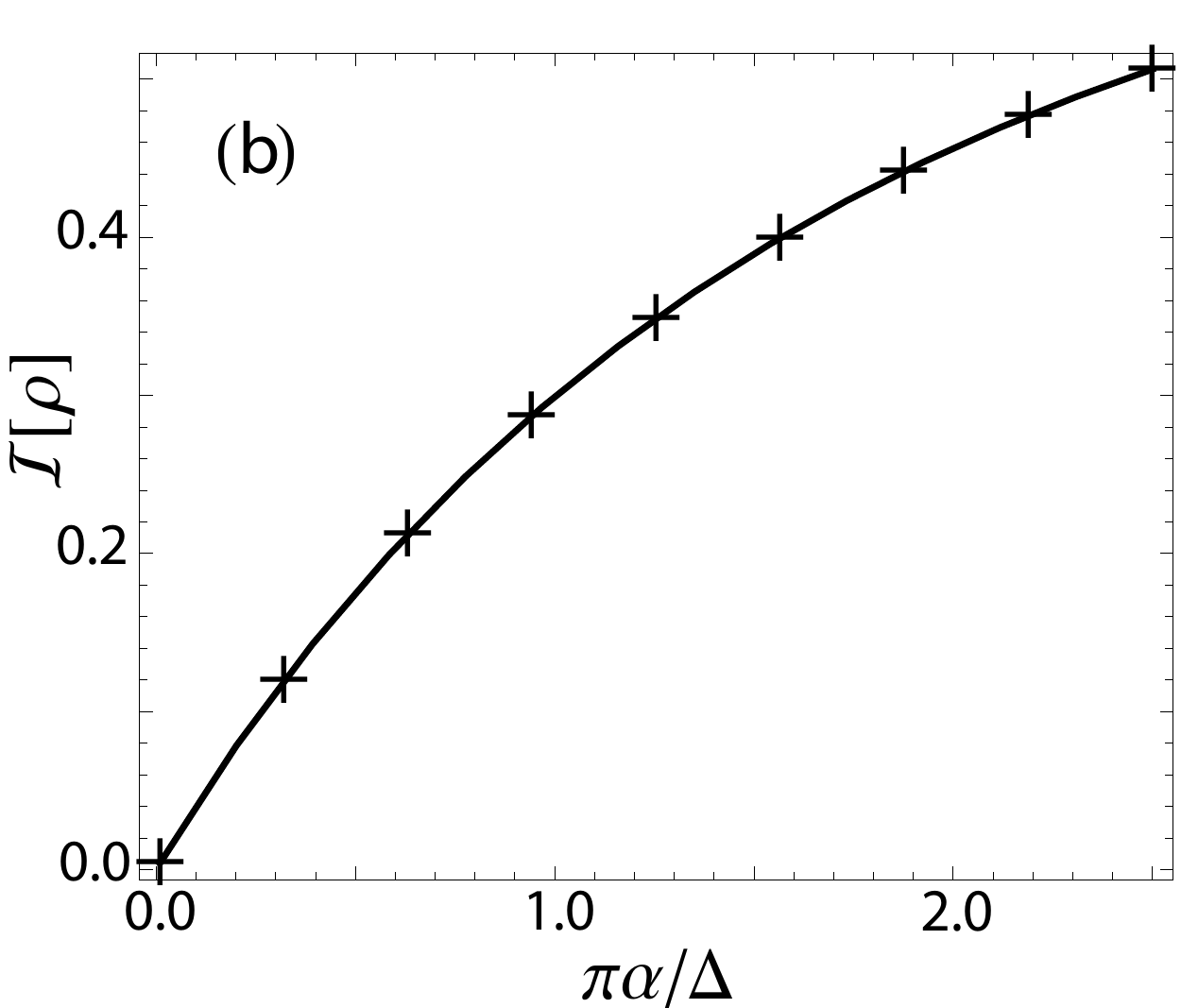}
\caption{(Color online) 
Steady-state non-Gaussianity (a) and
quantum mutual information 
(b) as a function of coupling strength, derived from the RCME (solid curve) and a thermal state with respect to the TLS-RC Hamiltonian $H_0$ (crosses), see Eq.~(\ref{eq:noncanonicalthermal}). Parameters: $\epsilon=0.5\Delta$,
$\omega_c=0.05\Delta$ and $\beta\Delta=0.95$ ($\approx 300$~K for $\Delta=200$~cm$^{-1}$).}
\label{fig:SS}
\end{figure}

Fig.~\ref{fig:NG} shows the 
dynamics of the non-Gaussianity and the QMI
at both strong and weak system-environment coupling $\alpha$.
One of the most striking features of these plots is the presence of two distinct timescales, which is most obvious for weak couplings, but is also present at stronger 
coupling strengths. At short times, the QMI is oscillatory, 
an indication of the 
memory effects 
implied by system-environment correlations, which also 
push the RC away from its initial Gaussian state.
At longer times, 
we see 
that system-environment correlations---and consequently non-Gaussian environmental states---are also generated
on a second 
timescale, 
and in fact persist into the steady state, with 
values of both the non-Gaussianity and QMI 
dependent on the coupling strength 
as demonstrated in Fig.~\ref{fig:SS}. 
We therefore find a situation in which the Born-Markov approximation breaks down on all timescales. 
It describes neither the short time transient dynamics, due to the
absence of bath memory effects in the Markov approximation, nor the long time 
behaviour, due to the generation of significant TLS-bath correlations.
This leads to non-Gaussian (and hence non-thermal) environmental states upon tracing out the TLS, neglected in the Born approximation.
Our line of enquiry also raises an intriguing
question: if correlations
can so dramatically affect
the Gaussian nature of the
environment, how do they
impact upon the state of the TLS?

\section{Noncanonical equilibrium states} 

\begin{figure}[t]
\center
\includegraphics[width=0.235\textwidth]{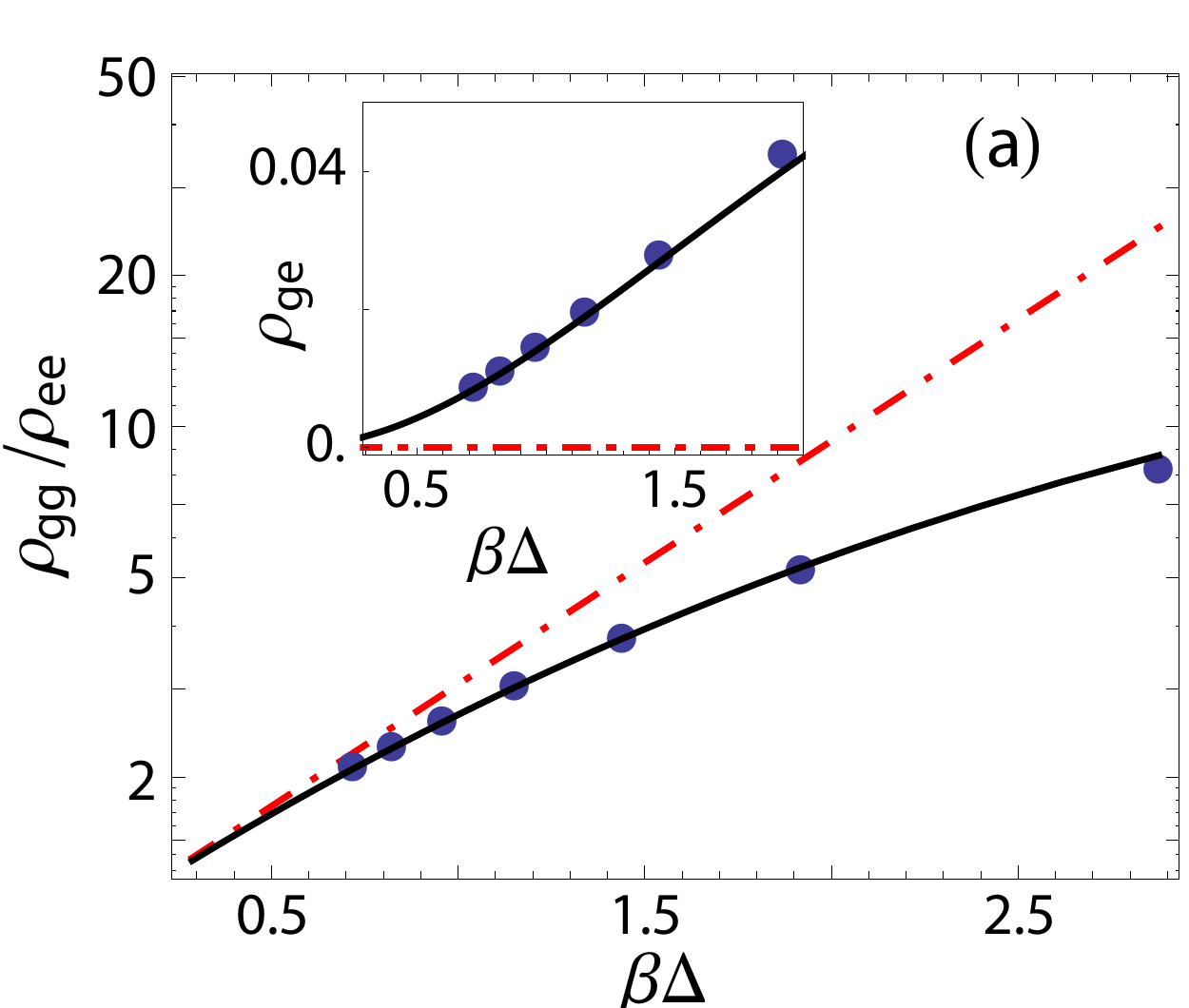}
\includegraphics[width=0.235\textwidth]{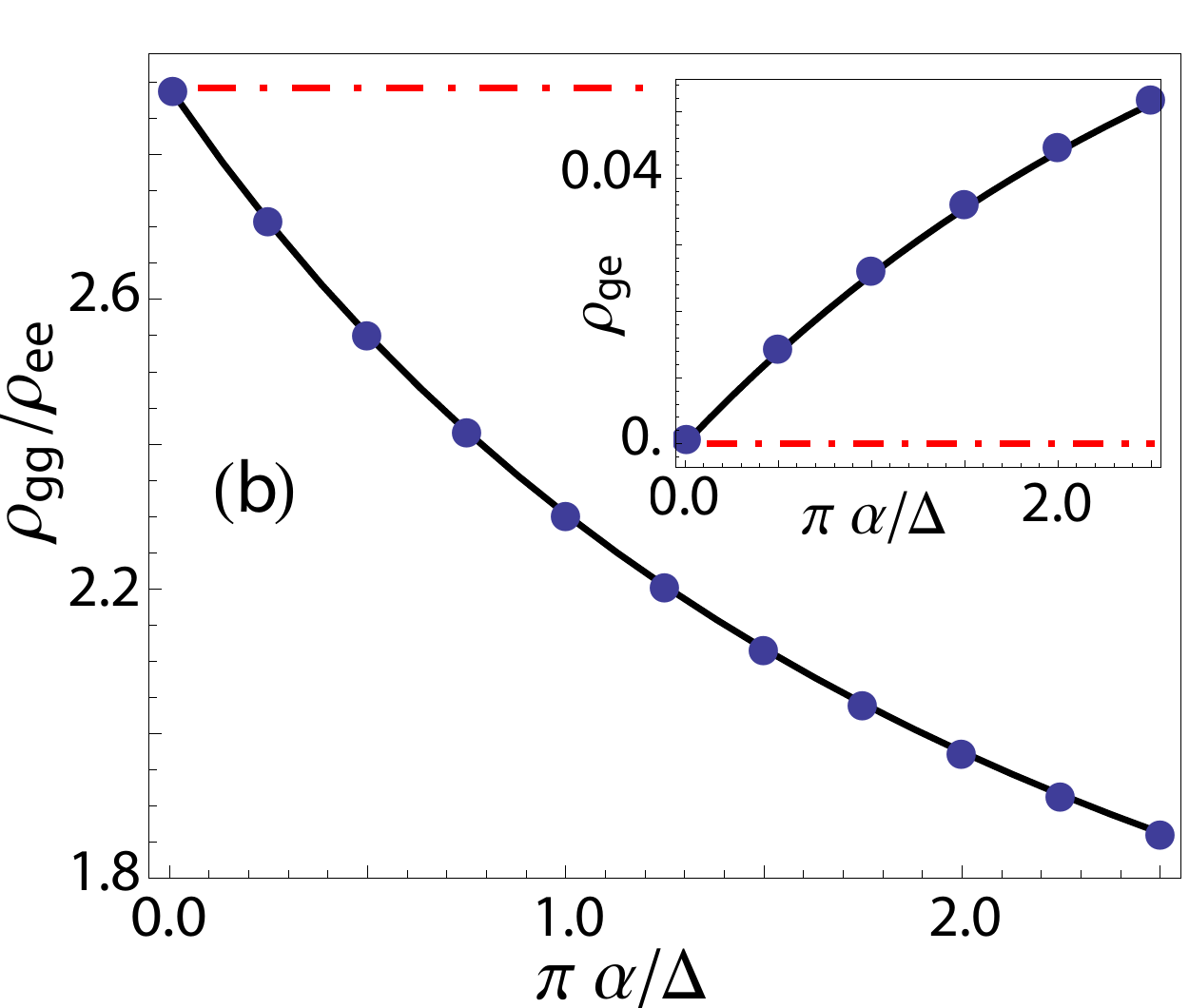}
\caption{(Color online) 
Steady state 
population ratio ($\rho_{\rm gg}/\rho_{\rm ee}$) and coherence ($\rho_{\rm ge}$) in the {\it eigenstate} basis of the TLS Hamiltonian $H_S$,
with $|{\rm g}\rangle$ ($|{\rm e}\rangle$) being the ground (excited) eigenstate: (a) against temperature (at $\pi\alpha=0.5\Delta$) and (b)
against coupling strength (at $\beta\Delta=0.95$). 
Shown are predictions from the HEOM (points), a thermal state with respect to the TLS-RC Hamiltonian $H_0$ (solid), and a thermal state with respect to
$H_S$ (dash-dot).
Parameters: $\epsilon=0.5\Delta$ and $\omega_c=0.05\Delta$. 
Considering $\Delta=200$~cm$^{-1}$ gives $T=575$~K for $\beta\Delta=0.5$, $T=288$~K for $\beta\Delta=1$, and $T=115$~K for $\beta\Delta=2.5$.}
\label{fig:steady}
\end{figure}

A clue can already be taken from Fig.~\ref{fig:SS}, where we find that
the steady-state properties of both the RC and TLS-RC correlations
may be 
described by a thermal state with respect to $H_0$, i.e.
\begin{align}\label{eq:noncanonicalthermal}
\rho_{\rm th}=\frac{e^{-\beta H_0}}{Z}=\frac{1}{Z}e^{-\beta\left(\frac{\epsilon}{2}\sigma_z+\frac{\Delta}{2}\sigma_x+\lambda\sigma_z(a^\dagger+a)+\Omega a^\dagger a\right)},
\end{align}
where $Z={\rm tr}(e^{-\beta H_0})$. 
Likewise, in Fig.~\ref{fig:steady} we show that the equilibrium
behaviour of the TLS (after tracing out the mode) clearly departs from the canonical statistics expected from a perturbative, weak coupling treatment of the environmental influence, which would be given instead by a thermal state with respect only to 
the TLS Hamiltonian, $H_S=\frac{\epsilon}{2}\sigma_z+\frac{\Delta}{2}\sigma_x$.

Fig.~\ref{fig:steady}(a) 
explores the temperature dependence of the TLS equilibrium state 
in the energy eigenbasis. To emphasise the departure 
from canonical statistics we have plotted the population ratio on a logarithmic scale.
For a canonical distribution we expect a linear dependence on the inverse temperature, shown by the dash-dot line, and given by $\ln({\rho_{gg}}/{\rho_{ee}})=-\beta\eta$, where $\eta=\sqrt{\epsilon^2+\Delta^2}$ is the TLS splitting. 
Crucially, however, the RCME steady state shows a clear deviation from this linear behaviour with decreasing temperature. To provide context, taking our estimate again of $\Delta=200$~cm$^{-1}$ relevant to molecular systems, we see that deviations from canonical statistics begin to become apparent around $\beta\Delta=0.95\approx300$~K, and should thus be observable {\it even at room temperature} in such systems. Note, from the inset, that a non-zero level of coherence is also apparent around such temperatures (and lower), 
again pointing to the noncanonical nature of the TLS equilibrium state. 
These departures demonstrate that we cannot represent the TLS steady state as a Gibbs distribution over the system eigenstates. They also quantify relatively simple experimental signatures of the breakdown of canonical statistics in open quantum systems. For example, by measuring only the TLS populations over a reasonable temperature range, we may infer the emergence of noncanonical equilibrium states simply by observing a nonlinear temperature dependence as shown. 

Finally, in Fig.~\ref{fig:steady} (b) we examine the
TLS steady state as a function of system-bath coupling strength at constant temperature. 
We again see significant deviations from the (unchanging) canonical thermal state, 
such as the development of steady state coherences within the TLS eigenstate basis, which are apparent for any non-vanishing coupling strength.
Nevertheless, 
it is evident 
from all plots in Fig.~\ref{fig:steady}, 
by the agreement between Eq.~(\ref{eq:noncanonicalthermal}) and the HEOM, that the noncanonical steady state is still extremely well described by a thermal state across a wide range of parameters, though now in the mapped TLS-RC representation. This behaviour is a clear result of the
TLS-environment correlations;
the TLS cannot
be considered merely as being in a product state with a (thermal) environment.
The non-separability of the steady state thus implies that
the equilibrium behaviour of the TLS cannot be described as a thermal distribution over its eigenstates either, but rather, one should also consider
states of the environment within the description.
This is markedly
different to standard master equation techniques and statistical mechanics approaches, where steady
states are commonly characterised by the bath temperature and
system Hamiltonian alone, an artefact of applying the Born
approximation.

\section{Summary}

To summarise, by exploiting a collective coordinate mapping, we have 
derived a
master equation 
valid in the non-adiabatic regime of the spin-boson model. 
Notably, besides an accurate description of the system dynamics, our approach also allows us to quantify the accumulation 
of system-environment correlations with time, 
as well as probe the dynamic evolution of states of the environment. We have shown that properly accounting for the generation of system-environment correlations is essential for describing both the transient dynamics and equilibrium distributions of open systems in this regime. 
In particular, we have demonstrated that long-lived correlations lead to the emergence of noncanonical system equilibrium states that can be characterised in a simple and
intuitive way within our formalism, as thermal states
of the system-collective coordinate
Hamiltonian. 
Our approach can be applied to a number of systems of
practical relevance.
For example, in molecular and solid-state (e.g.~superconducting) devices~\cite{leggett87,weissbook,breuer2007theory,lehur07,PhysRevB.85.140506,wurger98,mahan,goldstein13,nitzanbook,maykuhnbook,garg:4491,gilmore05,engel2007evidence,lee07,collini09sc,PhysRevB.87.075422,doi:10.1142/S0217979213450537,PhysRevLett.111.243602}, system-environment coupling can be
strong and 
memory effects important. 
Taking parameters relevant for energy transfer processes in molecular dimers, we have shown that 
deviations from canonical statistics could be observable even at room temperature, which raises the intriguing question of the role of noncanonical equilibrium states in larger molecular aggregates~\cite{engel2007evidence,lee07,collini09sc}. 
Finally, a proper understanding of 
equilibrium states 
is a vital component in the growing field exploring the thermodynamics of quantum systems; the full implications of noncanonical steady-states 
in open quantum systems~\cite{cklee12,martinez13}
thus constitutes a fascinating topic for future exploration.

\section{Acknowledgements} 

We would like to thank Alex Chin, Jianshu Cao, Javier Cerrillo, Dara McCutcheon, Tom Stace, and Tommaso Tufarelli for interesting discussions. J.~I.-S.~is supported by the EPSRC and A.~N.~by Imperial College and The University of Manchester. N.~L. acknowledges the hospitality of the Controlled Quantum Dynamics Group at Imperial College.

\appendix

\section{Reaction coordinate mapping}\label{RCmap}

As in the main text, we start by considering a two level system (TLS) coupled linearly to a harmonic environment described by the spin-boson (SB) Hamiltonian (with $\hbar=1$):
\begin{equation}\label{eq:SBH}
H=\frac{\epsilon}{2}\sigma_z+\frac{\Delta}{2}\sigma_x+\sigma_z\sum\limits_{k}f_k(c_k^\dagger+c_k)+\sum\limits_k\nu_kc^\dagger_kc_k,
\end{equation}
where $\sigma_i$ ($i=x,y,z$) are the standard TLS Pauli operators in a basis where $\sigma_z=\ket{1}\bra{1}-\ket{2}\bra{2}$, and $c_k$ ($c_k^\dagger$) are creation (annihilation) operators for bosonic modes of frequency $\nu_k$, which couple to the TLS with strength $f_k$. The system-environment coupling is fully specified by the spin-boson spectral density, $J_{\rm SB}(\omega)=\sum_kf_k^2\delta(\omega-\omega_k)$.

We now apply a normal mode transformation to Eq.~(\ref{eq:SBH}), leading to a mapped Hamiltonian of the form~\cite{martinazzo:011101}
\begin{align}\label{eq:RCH}
H_{\rm RC}&=\frac{\epsilon}{2}\sigma_z+\frac{\Delta}{2}\sigma_x+\lambda\sigma_z(a^\dagger+a)+\Omega a^\dagger a+\sum\limits_k\omega_kb^\dagger_kb_k\nonumber\\
&+(a^\dagger+a)\sum\limits_k g_k(b_k^\dagger+b_k)+(a^\dagger+a)^2\sum\limits_k\frac{g_k^2}{\omega_k},
\end{align}
where we have defined the collective (reaction) coordinate such that
\begin{equation}
\lambda(a^\dagger+a)=\sum\limits_kf_k(c_k^\dagger+c_k).
\end{equation}
The TLS-reaction coordinate (RC) coupling strength is given by $\lambda^{2}=\sum_kf_k^2$, such that the RC creation and annihilation operators satisfy the canonical commutation relation $[a,a^\dagger]=1$. This choice of coupling also fixes the RC frequency to be $\Omega^2=\lambda^{-2}\sum_k\nu_kf_k^2$. The RC is now coupled linearly to a residual harmonic environment characterised by the spectral density $J_{\rm RC}(\omega)=\sum_kg_k^2\delta(\omega-\omega_k)$. There is also a term quadratic in the system operators in Eq.~(\ref{eq:RCH}), known as the counter term, which is used to renormalise the mode frequency and avoid divergences due to friction~\cite{hartmann:11159}. 

In order to fully specify the mapping described above, we must relate the RC and SB spectral densities. We do this by following the procedure first outlined by Garg \emph{et al.}~\cite{garg:4491} 
and derive the spectral density, both before and after the mapping, from the classical equations of motion~\cite{PhysRevB.30.1208}.  
Since the spectral density doesn't contain any information about the system itself, but rather just the coupling strength between the system and environment, it then follows that in Eq.~(\ref{eq:SBH}) we can swap the TLS for a continuous classical coordinate $q$ moving in a potential $V(q)$. This yields a Hamiltonian of the form
\begin{align}~\label{eq:contH}
H_{q}&=\frac{P_q^2}{2}+U(q)+q\sum\limits_k \tilde{f}_k \hat{x}_k +q^2\sum\limits_k \frac{\tilde{f}_k^2}{2\nu_k^2}\nonumber\\
& +\frac{1}{2}\sum\limits_k \left(\hat{p}_k^2+\nu_k^2 \hat{x}_k^2\right),
\end{align}
where, for simplicity, we have written Eq.~(\ref{eq:contH}) in the position representation, with coordinate and momentum operators of the environment defined as
\begin{equation}
\hat{x}_k=\sqrt{\frac{1}{2\nu_k}}(c^\dagger_k+c_k)\hspace{10pt}\text{and}\hspace{10pt}\hat{p}_k=i\sqrt{\frac{\nu_k}{2}}(c^\dagger_k-c_k),
\end{equation}
and the coupling between the TLS and the $k^{th}$ mode of the environment is given by $\tilde{f}_k=\sqrt{2\nu_k}f_k$. From this Hamiltonian, we attain a set of classical equations of motion:
\begin{align}
\ddot{q}(t)&=-U^\prime(q)-\sum\limits_k \tilde{f}_k x_k(t) -q(t)\sum\limits_k \frac{\tilde{f}_k^2}{\nu_k^2},\nonumber\\
\ddot{x}_k(t)&=-\tilde{f}_k q(t)-\nu_k^2x_k(t).
\end{align}
We can eliminate the bath variables from the equation of motion for the classical coordinate by making use of the Fourier transform,
$\tilde{h}(z)=\int_{-\infty}^{\infty}h(t) e^{-iz t} dt$. This leads to an equation of the form
$\tilde{K}(z)\tilde{q}(z)=-\tilde{U}^\prime(q)$,
where the Fourier space operator is defined as
\begin{align}
\tilde{K}(z)&=-z^2\left(1+\sum\limits_k\frac{\tilde{f}_k^2}{\nu_k^2(\nu_k^2-z^2)}\right)\nonumber\\
&=-z^2\left(1+\int\limits_0^\infty d\nu \frac{J_{\rm SB}(\nu)}{\nu\left(\nu^2-z^2\right)}\right).
\end{align}
We solve the integral using the residue theorem, making use of a single pole at $\nu=z$, giving
$\tilde{K}(z)=-z^2+i\pi J_{\rm SB}(z)$. By writing $z=\omega-i\epsilon$, it follows that~\cite{PhysRevB.30.1208}
\begin{equation}\label{eq:SBcond}
J_{\rm SB}(\omega)=\frac{1}{\pi}\lim_{\epsilon\to 0+}{\rm Im}\left[K(\omega-i\epsilon)\right].
\end{equation}

We now use the same procedure to write $\tilde{K}(z)$ in terms of the RC spectral density. Since the mapping at this stage is exact, then the Fourier transformed operator, $\tilde{K}(z)$, will be identical before and after the normal mode transformation. If we once again swap the TLS for a continuous coordinate, and write the RC Hamiltonian given in Eq.~(\ref{eq:RCH}) in position space, then we have
\begin{align}
H_q&=\frac{P^2_q}{2}+U(q)+\kappa q\hat{x} +\frac{\kappa^2}{2\Omega^2}q^2+\frac{1}{2}\left(\hat{p}^2+\Omega^2\hat{x}^2\right)
\nonumber\\&+\hat{x}\sum\limits_{k}\tilde{g}_k\hat{X}_k+ \hat{x}^2\sum\limits_k\frac{\tilde{g}_k^2}{2\omega_k^2}+\frac{1}{2}\sum\limits_{k}\left(\hat{P}_k^2+\omega_k\hat{X}_k^2\right),
\end{align}
where we have scaled the coupling strengths such that $\kappa=\sqrt{2\Omega}\lambda$ and $\tilde{g}_k=\sqrt{2\omega_k}g_k$, and the position and momentum operators are defined in the usual way. This Hamiltonian leads to classical equations of motion of the form
\begin{align}
&\ddot{q}+\kappa \hat{x}+\frac{\kappa^2}{\Omega^2}q=-U^\prime(q),\nonumber\\
&\ddot{\hat{x}}+\left(\Omega^2+\sum\limits_k \frac{\tilde{g}_k^2}{\omega_k^2}\right)\hat{x}+\kappa^2 q+\sum\limits_k \tilde{g}_k \hat{X}_k=0,\nonumber\\
&\ddot{\hat{X}}_\alpha+\tilde{g}_\alpha\hat{x}+\omega^2_\alpha\hat{X}_\alpha=0.
\end{align}
By moving to Fourier space and eliminating both the RC and environment from the equation of motion for the classical coordinate, we get the expression for the Fourier space operator
$$\tilde{K}(z)=-z^2-\frac{\kappa^2}{\Omega^2}\frac{\mathcal{L}(z)}{\Omega^2+\mathcal{L}(z)},$$
where, by using the definition of the RC spectral density, we have
\begin{align}
\mathcal{L}(z)&=-z^2\left(1+\sum\limits_k\frac{\tilde{g}_k^2}{\omega_k^2(\omega_k^2-z^2)}\right)\nonumber\\&=-z^2\left(1+4\Omega\int\limits_{0}^{\infty}\frac{J_{\rm RC}(\omega)}{\omega \left(\omega^2-z^2\right)}d\omega\right).
\end{align}
Considering the RC to have an Ohmic spectral density of the form $J_{\rm RC}(\omega)=\gamma\omega\exp\left\{-\omega/\Lambda\right\}$, the integral reduces to
$\mathcal{L}(z)=-z^2+2i\pi\Omega\gamma z$, in the limit $\Lambda\rightarrow\infty$. By plugging this into Eq.~(\ref{eq:SBcond}), we get the condition
\begin{align}\label{eq:mappingSB}
J_{\rm SB}(\omega)&=\frac{1}{\pi}\lim_{\epsilon\to 0+}{\rm Im}\left[\tilde{K}(\omega-i\epsilon)\right]\nonumber\\&=\frac{1}{\pi}\frac{2\pi\Omega\gamma\kappa^2\omega}{(\Omega^2-\omega^2)^2+(2\pi\Omega\gamma)^2\omega^2}.
\end{align}
In order for Eq.~(\ref{eq:mappingSB}) to be consistent with the original spin-boson spectral density, we identify the relations
\begin{alignat}{2}
\omega_c=&\frac{\Omega}{2\pi\gamma}, &\qquad\text{and}\qquad\alpha=&\frac{2\lambda^2}{\pi\Omega},
\end{alignat}
which give us
\begin{equation}
J_{\rm SB}(\omega)=\frac{\alpha\omega_c\omega}{\omega^2_c+\frac{\omega^4\omega_c^2}{\Omega^2}+\omega^2}.
\end{equation}
Finally, by assuming $\omega_c\ll\Omega$, we get the spin-boson spectral density in Lorentz-Drude form,
\begin{equation}\label{eq:sbsd}
J_{\rm SB}(\omega)=\alpha\frac{\omega_c\omega}{\omega^2+\omega_c^2}.
\end{equation}

\section{Reaction coordinate master equation}
\label{sec:appendme}
We now wish to use Eq.~(\ref{eq:RCH}) to derive a master equation that treats the TLS-RC coupling exactly (up to some number of basis states) while the coupling to the residual environment is treated to second order (within a Born-Markov approximation). To this end, we define the interaction Hamiltonian as
\begin{equation}
H_I=\hat{A}\otimes \hat{B}+\tilde\lambda\ \hat{A}^2,
\end{equation}
where $\hat{A}=a^\dagger+a$, $\hat{B}=\sum_kg_k(b^\dagger_k+b_k)$ and $\tilde\lambda=\sum_kg^2_k/\omega_k$. To derive the RC master equation we first move into the interaction picture using the transformation $\tilde{H}_I(t)=\exp\{i(H_0+H_B)t\}H_I\exp\{-i(H_0+H_B)t\}=\hat{A}(t)\otimes \hat{B}(t)+\tilde\lambda\ \hat{A}^2(t)$, where $H_0=\frac{\epsilon}{2}\sigma_z+\frac{\Delta}{2}\sigma_x+\lambda\sigma_z(a^\dagger+a)+\Omega a^\dagger a$, is the TLS-RC Hamiltonian, and $H_B=\sum_k\omega_kb^\dagger_kb_k$. Tracing out the residual environment, we then find a 
master equation for the reduced TLS-RC density operator $\rho(t)$, which in the interaction picture may be written as
\begin{align}\label{eq:RCme1}
\frac{\partial\tilde\rho(t)}{\partial t}&=-i\ {\rm tr}_B\left[\tilde{H}_I(t),\rho(0)\otimes\rho_B\right]\nonumber\\&-\int\limits_0^t d\tau\ {\rm tr}_B\left[\tilde{H}_I(t),\left[\tilde{H}_I(t-\tau),\tilde\rho(\tau)\otimes\rho_B\right]\right],\nonumber\\\\
&=-i\ \tilde{\lambda}\left[\hat{A}^2(t),\rho(0)\right]\nonumber\\&-\tilde\lambda^2\int\limits_0^t d\tau\left[\hat{A}^2(t),\left[\hat{A}^2(t-\tau),\tilde\rho(\tau)\right]\right]\nonumber\\
&-\int\limits_0^t d\tau\bigg(
\left[\hat{A}(t),\left[\hat{A}(t-\tau),\tilde\rho(t)\right]\right]\Gamma^+(\tau)
\nonumber\\&+\left[\hat{A}(t),\left\{\hat{A}(t-\tau),\tilde\rho(\tau)\right\}\right]\Gamma^-(\tau)\bigg).
\end{align}
Here, we have made use of the Born approximation, that is, we have assumed that the system and residual environment remain in the product state $R(t)\approx\rho(t)\otimes\rho_B$ for all time, where $\rho_B=\exp(-\beta\sum_k\omega_k b_k^\dagger b_k)/{\rm tr}\{\exp(-\beta\sum_k\omega_k b_k^\dagger b_k)\}$. The correlation functions are defined as $\Gamma^\pm={\rm tr}\left\{(\hat{B}(\tau)\hat{B}\pm \hat{B}(-\tau)\hat{B})\rho_B\right\}/2$. In the continuum limit we may write them as
\begin{equation}
\Gamma^+(\tau)=\int\limits_0^\infty d\omega J_{\rm RC}(\omega)\coth\frac{\beta\omega}{2} \cos\omega\tau
\end{equation}
and
\begin{equation}
\Gamma^-(\tau)=i\int\limits_0^\infty d\omega J_{\rm RC}(\omega)\sin\omega\tau.
\end{equation}
As outlined in the previous section, the RC spectral density takes the form $J_{\rm RC}(\omega)=\gamma\omega \exp(-\omega/\Lambda)$, in the limit that that the cut-off frequency $\Lambda\longrightarrow\infty$. We can simplify Eq.~(\ref{eq:RCme1}) further by noticing that
\begin{align}
-i\tilde\lambda\left[\hat{A}^2(t),\tilde\rho(t)\right]&=-i\tilde{\lambda}\left[\hat{A}^2(t),\tilde\rho(0)\right]\nonumber\\&-\tilde\lambda^2\int\limits_0^t d\tau\left[\hat{A}^2(t),\left[\hat{A}^2(t-\tau),\tilde\rho(\tau)\right]\right].
\end{align}
By substituting this into the above master equation and assuming a Markov limit, that is, we take 
the time integrals to infinity, we acquire the RC master equation
\begin{align}\label{eq:me1}
\frac{\partial\tilde\rho(t)}{\partial t}&=-i\tilde\lambda\left[\hat{A}^2(t),\tilde\rho(t)\right]\nonumber\\&-\int\limits_0^\infty \int\limits_0^\infty d\tau d\omega J_{\rm RC}(\omega)\coth\frac{\beta\omega}{2} \cos\omega\tau\nonumber\\&\;\;\;\;\;\;\;\;\;\;\;\;\times\left[\hat{A}(t),\left[\hat{A}(t-\tau),\tilde\rho(t)\right]\right]\nonumber\\
&-i\int\limits_0^\infty\int\limits_0^\infty d\tau d\omega J_{\rm RC}(\omega)\sin\omega\tau\nonumber\\&\;\;\;\;\;\;\;\;\;\;\;\;\times\left[\hat{A}(t),\left\{\hat{A}(t-\tau),\tilde\rho(t)\right\}\right].
\end{align}
The definition of $J_{\rm RC}(\omega)$ assumes an infinite cut-off frequency, hence the first and last terms of Eq.~(\ref{eq:me1}) are divergent. However, we can eliminate the divergent contributions by integrating the last term by parts, such that
\begin{align}
\int\limits_0^\infty d\tau \sin\omega\tau\hat{A}(t-\tau)&=-\mathcal{P}\left(\frac{\hat{A}(t)}{\omega}\right)\nonumber\\&+\int\limits_0^\infty d\tau\frac{\cos\omega\tau}{\omega}\frac{\partial \hat{A}(t-\tau)}{\partial \tau}.
\end{align}
Using the principal value part of this integral we can cancel the counter term, which, after moving back into the Schr\"odinger picture, gives the master equation
\begin{align}\label{eq:master}
\frac{\partial\rho(t)}{\partial t}&=-i\left[H_0,\rho(t)\right]\nonumber\\&-\int\limits_0^\infty\int\limits_0^\infty d\tau d\omega J_{\rm RC}(\omega)\cos\omega\tau\coth\frac{\beta\omega}{2}\nonumber\\&\;\;\;\;\;\;\;\;\;\;\;\;\times\left[\hat{A},\left[\hat{A}(-\tau),\rho(t)\right]\right]\nonumber\\
&-\int\limits_0^\infty\int\limits_0^\infty d\tau d\omega J_{\rm RC}(\omega)\frac{\cos\omega\tau}{\omega}\nonumber\\&\;\;\;\;\;\;\;\;\;\;\;\;\times\left[\hat{A},\left\{\left[\hat{A}(-\tau),H_0\right],\rho(t)\right\}\right].
\end{align}

In order to derive the interaction picture system operators we shall diagonalise the TLS-RC system Hamiltonian, $H_0$, numerically. Let $\ket{\varphi_n}$ be an eigenstate of the system Hamiltonian, such that
$
H_0\ket{\varphi_n}=\varphi_n\ket{\varphi_n}.
$
We can now write the position operators in this eigenbasis:
\begin{equation}
\hat{A}=\sum\limits_{jk}A_{j,k}\ket{\varphi_j}\bra{\varphi_k},
\end{equation}
where $A_{j,k}=\langle \varphi_j\vert\hat{A}\vert\varphi_k\rangle$. In the interaction picture this becomes
\begin{equation}
\hat{A}(t)=\sum\limits_{jk}A_{j,k}e^{i\xi_{jk}t}\ket{\varphi_j}\bra{\varphi_k},
\end{equation}
where $\xi_{nm}=\varphi_n-\varphi_m$ is the difference between the $n^{th}$ and $m^{th}$ eigenvalues. Using this definition we can include the rates from Eq.~(\ref{eq:master}) into the operators, such that
\begin{align}
\hat{\chi}&=\int\limits_0^\infty\int\limits_0^\infty d\omega d\tau\ J_{\rm RC}(\omega)\cos\omega\tau\coth\frac{\beta\omega}{2}\hat{A}(-\tau)\nonumber\\&
\approx\frac{\pi}{2}\sum\limits_{jk}J_{\rm RC}(\xi_{jk})\coth\frac{\beta\xi_{jk}}{2}A_{jk}\ket{\varphi_j}\bra{\varphi_k},\\
\hat{\Xi}&=\int\limits_0^\infty\int\limits_0^\infty d\omega d\tau\frac{J_{\rm RC}(\omega)\cos\omega\tau}{\omega}\left[H_0,\hat{A}(-\tau)\right]\nonumber\\&
\approx\frac{\pi}{2}\sum\limits_{jk}J_{\rm RC}(\xi_{jk})A_{jk}\ket{\varphi_j}\bra{\varphi_k},
\end{align}
where we have neglected the imaginary Lamb shift terms. We can now write the master equation as
\begin{equation}
\frac{\partial\rho(t)}{\partial t}=-i\left[H_0,\rho(t)\right]-\left[\hat{A},\left[\hat{\chi},\rho(t)\right]\right]+\left[\hat{A},\left\{\hat{\Xi},\rho(t)\right\}\right].
\end{equation}

\section{HEOM}\label{heom}

The hierarchical equations of motion (HEOM) are a set of time-local equations
for the reduced system dynamics, governed by the spin-boson Hamiltonian [Eq.~(\ref{eq:SBH})], which capture the bath
dynamics and system-bath correlations through a set of auxiliary
density matrices. These equations are exact under the assumption
of a Lorentz-Drude spectral density, as given in Eq.~(\ref{eq:sbsd}), and an initially separable system-bath
state at $t=0$. Here we assume only a single bath such that the HEOM
can be written as \beq
\dot{\rho}_{\mathbf{n}} &=& -\left(i[H_S,\rho_{\mathbf{n}}] +
\sum_{m=0}^K \mathbf{n}_{m} \mu_m\right ) \rho_{\mathbf{n}} -
i\sum_{m=0}^K\left[Q,\rho_{\mathbf{n}_{m}^+}\right]\nonumber\\
&-& i\sum_{m=0}^K n_{m}\left(c_mQ\rho_{\mathbf{n}_{m}^-} - c_m^*
\rho_{\mathbf{n}_{m}^-}Q\right)\\
&-& \left(\frac{\alpha_H}{\beta \omega_c} - i \frac{\alpha_H}{2}
-\sum_{m=0}^K\frac{c_k}{\mu_k}\left[Q,\left[Q,\rho_{\mathbf{n}_{m}}\right]\right]\right),\nonumber
\eeq  where $Q=\sigma_z$ is the system-environment coupling operator, $\alpha_H=\pi \alpha$, and $H_S=\frac{\epsilon}{2}\sigma_z+\frac{\Delta}{2}\sigma_x$.
The bath correlation functions for the Lorentz-Drude spectral density are
\beq C = \sum_{m=0}^{\infty} c_{m} \exp\left(-\mu_{m}
t\right).\eeq Here $\mu_{0} = \omega_c$, $\mu_{m\geq 1} = 2\pi
m/\beta$, and the coefficients are \beq c_{0} = \frac{\omega_c
\alpha_H}{2}\left[\cot(\beta  \omega_c/2) - i\right]/\hbar\eeq and
\beq c_{m\geq 1} = \frac{2\alpha_H \omega_c}{\beta}
\frac{\mu_{m}}{\mu_{m}^2 -\omega_c^2}. \eeq Each density matrix is
labelled by  an index  of positive integers $\mathbf{n}$. As here
we only have a single bath the integers are defined as
$\mathbf{n}=\{n_1,n_2,...,n_m,...,n_K\}$. For each ``Matsubara''
term $m$ each index runs from $0$ to $\infty$. The null label
$\mathbf{n}=0=\{0,0,0....\}$ defines the system density matrix,
and any non-zero label refers to an auxiliary density matrix which
encodes the correlations with the bath. The terms in the equation
of motion $\mathbf{n}_{m}^{\pm}$ refer to an increase or decrease
by $1$ of the label index $m$. We take a cut-off in the
overall tier $N_c = \sum_{m} n_{m}$ and in the Matsubara terms $K$
which give convergence in the numerical results.

\section{Quantum mutual information and non-Gaussianity}\label{QMInonGauss}
In this Appendix we show that the mutual information for the TLS and RC acts as a lower bound to the correlations shared between the system and the original multi-mode environment.

Let $\chi$ be the density matrix describing the state of both the system and environment in the original spin-boson representation. We also define the reduced states of the system, $\rho_s={\rm tr}_E \chi$, and the environment, $\rho_E={\rm tr}_{s}\chi$. Let $\mathcal{U}=\openone_S\otimes\mathcal{R}$ be the unitary transformation that maps the spin-boson model to the RC model. Applying this unitary to the reduced state of the system has no effect due to the trace over the environment. However, on the reduced state of the environment this unitary transforms the spin-boson basis to that of the RC and residual bath, that is, $\mathcal{R}^\dagger\rho_{E}\mathcal{R}=\rho_{RC+B}$. 

The quantum mutual information for the system and the spin-boson environment is given by:
\begin{equation}\label{eq:appsbmutinfo}
\begin{split}
\mathcal{I}\left(\rho_s:\rho_E\right)=&S(\rho_s)+S(\rho_{E})-S(\chi),\\
=&S(\rho_s)+S(\rho_{RC+B})-S(\tilde{\chi}),\\
\end{split}
\end{equation}
where $S(\varrho)=-tr\left(\varrho\ln\varrho\right)$ is the von-Neumann entropy and $\tilde{\chi}=\mathcal{U}^\dagger\chi\mathcal{U}$ is the total state in the RC basis. Here we have used the unitary equivalence of the von-Neumann entropy to write the quantum mutual information in terms of the RC basis.

To proceed we shall make use of the strong subadditivity of the von-Neumann entropy, that is:
\begin{equation}
S(\tilde\chi)+S(\rho_{RC})\le S(\rho_{S+RC})+S(\rho_{RC+B}),
\end{equation}
where $\rho_{S+RC}={\rm tr}_{B}\tilde{\chi}$, with the trace taken over the residual environment. Using this property in conjunction with Eq.~(\ref{eq:appsbmutinfo}) gives:
\begin{align}
\mathcal{I}\left(\rho_s:\rho_E\right)&\ge S(\rho_S)+S(\rho_{RC+B})+S(\rho_{RC})-S(\rho_{s+RC})\nonumber\\&-S(\rho_{RC+B}),\nonumber\\
&\ge S(\rho_s)+S(\rho_{RC})-S(\rho_{S+RC}).
\end{align}
Therefore, we have the condition
\begin{equation}\label{eq:mutinn}
\mathcal{I}\left(\rho_s:\rho_E\right)\ge\mathcal{I}\left(\rho_s:\rho_{RC}\right).
\end{equation}
Hence, the mutual information between the system and RC acts as a lower bound to the mutual information between the system and spin-boson environment. 
Furthermore, in the limit that the Born approximation holds between the composite system  (TLS and RC) and the residual environment, that is $\tilde\chi\approx\rho_{s+RC}\otimes\rho_{B}$, then the inequality in Eq.~(\ref{eq:mutinn}) becomes an equality due to the additive nature of the von-Neumann entropy.

Similarly, the non-Gaussianity can be shown to be invariant under symplectic transformation (i.e. operators that are quadratic in field operators) and monotonically decreases under partial trace~\cite{PhysRevA.78.060303}. This means that the non-Gaussianity of the RC acts as a rigorous lower bound for the non-Gaussianity of the original spin-boson environment, that is:
\begin{equation}
\delta_G\left[\rho_E\right]\ge\delta_G\left[\rho_{RC}\right]. 
\end{equation}

\providecommand{\noopsort}[1]{}\providecommand{\singleletter}[1]{#1}%

\end{document}